\documentclass[a4paper]{article}
\usepackage{amsmath}
\usepackage{amssymb}
\newcommand{\haak}[1]{\left(#1\right)}

\newcommand{\lhaak}[1]{\left | #1\right |}

\newcommand{\gem}[1]{\left\langle #1\right\rangle}

\begin{document} 
\vskip 2cm
\title{Detecting Dark Matter using Centrifuging Techniques} 
\vskip 1.5cm 
\author{S. Mitra\\
{\it saibalm@science.uva.nl}\\
Instituut voor Theoretische Fysica\\
Universiteit van Amsterdam\\
1018 XE Amsterdam\\
The Netherlands\\
\and
R. Foot\\
{\it foot@physics.unimelb.edu.au}\\
School of Physics\\
Research Centre for High Energy Physics\\
The University of Melbourne\\
Victoria 3010 Australia} 
\date{January 2003}
\maketitle 
\begin{abstract}
A new and inexpensive technique for detecting self interacting dark matter 
in the form of small grains in bulk matter is proposed. 
Depending on the interactions with ordinary matter,
dark matter grains in bulk matter may be isolated by using a 
centrifuge and using ordinary matter as a filter. The 
case of mirror matter interacting with ordinary matter via 
photon-mirror photon kinetic mixing 
provides a concrete example of this type of dark matter candidate.
\end{abstract} 
\newpage 
It is known that a large fraction of the mass 
of the universe is in the form of dark matter. Most of this dark matter is believed to exist in the form of as of yet unknown elementary particles. Many different types of candidates have been proposed, such as weakly interacting massive particles (WIMPS), strongly interacting massive particles (SIMPS) and charged massive particles (CHAMPS). Despite many experimental searches all attempts to detect these particles have failed. For a review see\cite{perl}. 

Interestingly there is one possible dark matter candidate which has not
yet been experimentally scrutinized. The idea is that dark matter
particles may have strong enough 
self interactions such that they can condense into small grains, and
also
interact with ordinary matter, such that a grain can remain on the 
surface of the Earth. 
A specific 
candidate for this kind of dark matter is provided by theories
respecting mirror symmetry, as we will now briefly explain. 

Mirror symmetry appears broken by the interactions of the known elementary
particles (because of their left-handed weak interactions).
Nevertheless,
mirror symmetry can exist if one introduces for every particle a 
corresponding mirror particle, of exactly the
same mass as the ordinary particle
\cite{ly,flv}.
These mirror particles interact with each other in exactly the 
same way that the ordinary particles do. The mirror particles are not 
produced
(significantly) in laboratory experiments just because they couple
very weakly to the ordinary particles. In the modern language of
gauge theories, the mirror particles are all singlets under the
standard $G \equiv SU(3)\otimes SU(2)_L \otimes U(1)_Y$ gauge
interactions. Instead the mirror fermions interact with a set of
mirror gauge particles, so that the gauge symmetry of the theory
is doubled, i.e. $G \otimes G$ (the ordinary particles are, of
course, singlets under the mirror gauge symmetry)\cite{flv}.
Mirror symmetry is conserved because the mirror fermions experience $V+A$
(right-handed) mirror weak interactions and the ordinary fermions
experience the usual $V-A$ (left-handed) weak interactions.
Ordinary and mirror particles interact with each other
predominately by gravity (and possibly by new interactions
as we will explain below). 
Clearly, mirror matter is an ideal candidate for the
dark matter inferred to exist in the Universe because
it is dark and stable\cite{blin}. It also appears
to have the right properties to explain a number
of other interesting puzzles. 
For a review, see Ref.\cite{comet}. 

While we know that ordinary and mirror
matter do not interact with each other via any
of the {\it known} non-gravitational forces,
it is possible that new interactions exist which
couple the two sectors together.
In Ref.\cite{flv,flv2}, all such interactions consistent
with gauge invariance, mirror symmetry and renormalizability
were identified. Of most importance for this paper is
the photon-mirror photon kinetic mixing interaction. 
In quantum field theory,
photon-mirror photon kinetic mixing
is described by the interaction
\begin{equation}
{\cal L} = \frac{\epsilon}{2}F^{\mu \nu} F'_{\mu \nu},
\label{ek}
\end{equation}
where $F^{\mu \nu}$ ($F'_{\mu \nu}$) is the field strength tensor
for electromagnetism (mirror electromagnetism). This type of
Lagrangian term is gauge invariant and renormalizable and can
exist at tree level\cite{flv,fh} or may be induced radiatively in
models without $U(1)$ gauge symmetries (such as grand unified
theories)\cite{gl,bob,cf}. One effect of ordinary photon-mirror
photon kinetic mixing is to give the mirror charged particles a
small electric charge\cite{flv,gl,bob}. That is, they couple to
ordinary photons with electric charge $\epsilon e$. The most important 
experimental particle physics implication of photon-mirror
photon kinetic mixing is that it modifies the properties of 
orthopositronium\cite{gl}. The current experimental
situation is summarized in Ref.\cite{fg}, which
shows that $\lhaak{\epsilon} \lesssim 10^{-6}$,
with some evidence for $\lhaak{\epsilon}\approx 10^{-6}$ from
the 1990 vacuum cavity experiment\cite{vac}. 

Understanding the possible astrophysical implications of
photon-mirror photon kinetic mixing has been
the subject of a number of recent papers\cite{fiv,footyoon,fvpioneer,
eros}.
The existence of photon-mirror photon
kinetic mixing allows mirror matter to explain a number
of puzzling observations, including
the pioneer spacecraft anomaly\cite{study,fvpioneer}, 
anomalous meteorite events\cite{anom,footyoon}
and the unexpectedly low number of small craters on the asteroid
433 Eros\cite{near1,eros}. In Ref.\cite{eros}, it was shown that
these explanations require 
$\lhaak{\epsilon} \gtrsim 10^{-9}$. 
Thus, the most interesting parameter range for $\epsilon$ suggested
by observations is
\begin{equation}
10^{-9} \stackrel{<}{\sim} |\epsilon | \stackrel{<}{\sim} 10^{-6}.
\end{equation}

One other, perhaps very important
implication of photon-mirror photon kinetic mixing which we
have yet to mention is that it can provide a force which opposes
the effect of gravity, so that a mirror matter fragment can 
potentially remain on the Earth's surface.
Whether this actually happens, depends on the strength of the 
photon-mirror photon 
kinetic mixing compared to the weight of the fragment. 
If the mirror fragment is embedded inside ordinary matter, then the mirror 
atoms will have an average electrostatic energy induced by the 
photon-mirror photon kinetic mixing. The fragment can experience a 
strong force when this energy changes rapidly as a function of its 
position, e.g. at the boundary between a low and high density medium. 
In the appendix we derive the following equation for the 
electrostatic force exerted on a stationary fragment at the boundary between 
two media compared to its weight: 
\begin{equation}
\label{mon}
\frac{\lhaak{F_{\text{static}}}}{\lhaak{F_{\text{gravity}}}}
\sim \lhaak{\epsilon}10^{10} (\text{cm}/R)
\end{equation} 
Here, R is the size of the fragment. For positive $\epsilon$, the electrostatic force is (typically) directed from the 
high density medium toward the low density medium, 
while for negative $\epsilon$ the electrostatic force has 
the opposite direction. 
According to the above equation, a mirror 
matter fragment of size $R = 1\text{ cm}$ could
remain on the Earth's surface if $\epsilon$ is positive and
$\epsilon \gtrsim 10^{-10}$. [Of course, if it impacted with high
velocity, it would be buried some distance below the surface, as
we will discuss]. 
For $\epsilon$ less than $10^{-10}$
it would fall toward the center of the Earth.
If $\epsilon < 0$, then the mirror matter
fragment would necessarily move into the ground (because
in this case
the electrostatic force is then attractive between
the low density air and high density ground).
But, because the ground is of varying composition,
a fragment would stop after becoming completely 
embedded within the ground. The limit
for this to happen would be of the
same order of magnitude, i.e. $|\epsilon |
\gtrsim 10^{-10}$. 

If dark matter exists in the form of small grains in the ground, 
then one may try to isolate it by centrifuging soil samples. 
Modern ultracentrifuges are capable of an acceleration 
of about $10^{6} g$ . 
In the case of mirror matter Eq. (\ref{mon}) shows that increasing 
the force of gravity by a factor of a million will remove 
mirror matter fragments greater than 100 microns in size for 
$10^{-10} \lesssim |\epsilon| \lesssim 10^{-6}$. 
The simplest technique to detect the presence of small grains of 
dark matter is to first weigh a soil sample, then centrifuge it 
for some time\footnote{
The centrifugation time required can be estimated 
from $F_{\text{friction}}$,
Eq.(\ref{friction}), derived in the appendix.
The velocity of the fragment, relative to the spinning test
tube in a centrifuge, can be found by equating $F_{\text{friction}}
\approx F_{\text{acceleration}} \sim 10^6 g$. From
Eq.(\ref{friction}), this suggests $U \stackrel{>}{\sim} 
0.2$ cm/s. Taking into account possible uncertainties in our
approximations, a centrifugation time of 10 minutes should
be adequate.}, and then weigh it again. If there were indeed 
mirror matter grains present, then these should have been 
removed, leading to a lowering of the weight. In practice the 
sensitivity of such tests is limited to about one 
part in $10^6$ by weight (for a 100 gram sample). 
Still, nobody has ever done this type of experiment before. Such a 
sensitivity may well be enough to discover this type of dark matter, if
it exists (especially if the sample to be tested is chosen
appropriately, see the discussion below). 

A more sensitive test may be performed by attempting to catch 
the escaping dark matter fragments in a backing around 
the inside of the centrifuge. The backing should preferably be of inhomogeneous 
composition to maximize the probability of catching fragments in 
it. By centrifuging many soil samples the backing may 
become enriched with dark matter fragments. A direct weight 
measurement of the backing could confirm this. Alternatively, 
a sample of the backing may be centrifuged and tested for 
a decrease in weight as described above. Such a technique could 
yield a sensitivity of about
1 part in $10^7 - 10^8$ by weight. 
One might worry that the velocity of the fragments could be
too
large to be caught in the backing, however it turns out
that the frictional force of a mirror fragment moving
in ordinary matter is quite large, as we also show in
the appendix. 
The conclusion is that a mirror fragment
(with initial velocity $U_i$)
will slow down enough to enable it to be captured in
ordinary matter (of mass density $\rho$) after a distance of order:
\begin{eqnarray}\label{nice}
L &\sim & 
\frac{10^{-7}}{\Lambda}\haak{\frac{U_i}{300 \text{ m/s}}}\left(
\frac{4 \text{ g}/\text{cm}^3}{\rho}\right)
\ \text{ meters,}
\text{ for} \ U_i \lesssim 300 \text{ m/s} \nonumber \\
L &\sim & 
10^{-7}\left( \frac{U_i}{300 \text{ m/s}} \right)^4 \left( \frac{10^{-8}}
{\epsilon}\right)^2 \left( \frac{4 \text{ g}/\text{cm}^3}{\rho}\right)
\text{ meters,}
\text{ for} \ U_i \gtrsim 300 \text{ m/s}
\nonumber \\
\end{eqnarray}
where $\Lambda = 
\haak{|\epsilon |/10^{-8}}^{2}
$ for $|\epsilon | \lesssim 10^{-8}$ and $\Lambda = 1$ for
$|\epsilon| \gtrsim 10^{-8}$. 
Since the speed at which a fragment will leave the
centrifuge is less than about 1000 m/s, the above
equation suggests that a 
backing thickness greater than about a millimetre will be
adequate to capture small fragments. 

Besides weight measurements, there could be other ways to 
detect small dark matter grains escaping from a centrifuge. 
Since the interaction between the dark matter particles and ordinary 
atoms are strong enough to keep small grains from sinking into the 
Earth, these interactions may also be strong enough to cause 
a dark matter grain to thermalize with its environment on not 
too long time scales. The escaping dark matter grains may thus also 
be detected using cryogenic calorimeters. 

The next issue is what type of sample to use. If this
type of dark matter were present during the Earth's formation
it would be expected to be most abundant in the Earth's core.
However, such dark matter may also be extraterrestrial
in origin, for example it may come from small mirror
matter space bodies if they collide with the Earth.
In this case, this dark matter may be present on (or near)
the Earth's surface; enhanced
at various `impact sites'. 
Various candidate sites have been discussed in Ref.\cite{footyoon},
including Tunguska and a small yet specific site in Jordan.
Furthermore, according to Eq.(\ref{nice}) the
mirror matter fragments will be very close (centimeters!) to the surface
(since the impact velocity in both of these events is expected
to be less than 1 km/s because of atmospheric effects).
More generally, it has been known for a long time
that deep sea sediment is one
place where extraterrestrial material accumulates significantly.
It should also be a good place to
test for the existence of mirror matter-type dark matter. 

In conclusion, 
we have explored the possibility that dark matter may potentially 
exist on (or near) the Earth's surface. A specific example of
such dark matter is provided by mirror matter with
photon-mirror photon kinetic mixing interaction. This type
of dark matter has yet to be experimentally tested. We have 
therefore proposed
a new and inexpensive technique to directly test samples for 
the presence of this type of dark matter.
In the case of mirror matter, we have shown that this 
test is effective for mirror 
matter fragments larger than 100 microns in the range 
of $10^{-10} \lesssim |\epsilon| \lesssim 10^{-6}$, with
a sensitivity of up to 1 part in $10^8$. 
\vskip 0.5cm
\noindent
{\bf \large Acknowledgments}
\vskip 0.3cm
\noindent
R.F. would like to thank Peter Haines for some valuable discussions. 
\vskip 0.4cm 
\noindent
{\bf \large Appendix: Can a Mirror Matter Fragment Remain at the Earth's Surface?} 
\vskip 0.3cm 
In this appendix we will estimate the 
force on a mirror matter
fragment embedded in an ordinary
matter medium which is due to the photon-mirror photon
interaction. We will call this force $F^{\epsilon}$.
In general, for a fragment in motion
(with velocity $U$),
$F^{\epsilon}$ will contain a velocity dependent frictional term as well
as a static term,
that is
\begin{eqnarray}
F^{\epsilon} = F_{\text{static}} + F_{\text{friction}}
\label{aa1}
\end{eqnarray}
where $F_{\text{static}}$ is independent of $U$ 
and $F_{\text{friction}} \to 0$ as $U \to 0$.
We will first estimate $F_{\text{static}}$ and then consider
$F_{\text{friction}}$. 

Consider a mirror matter fragment with mass density $\rho'$, composed
of mirror atoms of mass $M_{A'}$, embedded within ordinary matter. 
Suppose this fragment is at the interface
of two homogeneous (ordinary matter) mediums, which we label
medium 1 and medium 2 (e.g. air/earth or earth/quartz etc). 
Let $A$ be the cross sectional area 
of the fragment
measured parallel to the interface. If the fragment moves a distance $dr$ orthogonal to the interface, 
then the number of mirror atoms moving from medium 1 to medium 2 is
simply $A\rho' dr/M_{A'}$ and the
electrostatic potential energy of the fragment will 
change by an amount $dE$:
\begin{equation}
dE=\haak{\gem{\zeta_{2}}-\gem{\zeta_{1}}}\frac{A\rho' dr}{M_{A'}}
\end{equation}
where $\gem{\zeta_{1}}$ ($\gem{\zeta_{2}}$) are the mean electrostatic 
energies coming from the interactions of mirror
atoms with the ordinary atoms of medium 1 (medium 2). Therefore it will experience an electrostatic force of: 
\begin{equation}\label{1}
F_{\text{static}}=\haak{\gem{\zeta_{1}}-\gem{\zeta_{2}}}\frac{A\rho'}{M_{A'}}\hat{n}
\end{equation} 
Here $\hat{n}$ is the unit normal vector of the interface, pointing from medium 1 to 2. 

The energies $\zeta_{1,2}$ are
very small because they are suppressed by
$|\epsilon | \lesssim 10^{-6}$ and 
are most significant 
when the mirror and ordinary nuclei are
close enough so that the screening effects
of the electrons can be approximately ignored.
If $z$ is the distance between the mirror nuclei and the
nearest ordinary nuclei, then
\begin{eqnarray}
\zeta (z) \simeq &\frac{ZZ' e^2 \epsilon}{z}
&\text{for } z \lesssim r_2\nonumber\\
\zeta (z)
\simeq &0 &\text{for } z \gtrsim r_2
\end{eqnarray}
where $Z$ is the atomic number of the ordinary atoms,
$Z'$ is the (mirror) atomic number of the mirror atoms.
The distance
$r_2$ is the radius over which significant
electrostatic interaction occurs, which we will
approximate to
the second Bohr radius, i.e.
\begin{equation}
r_2 \approx 4a_0/Z' \sim 10^{-9}\ \text{cm}, 
\label{bohr}
\end{equation}
where $a_0$ is the 
hydrogen Bohr radius. 

Because of rapid thermal motion and
the (typically) different
chemical composition and structure of the mirror
matter fragment and ordinary matter medium, to a good approximation,
the mean value of $\zeta_i$, $\langle \zeta_i \rangle$,
is simply the value of $\zeta (z)$ averaged over
the volume occupied by atoms, that is:
\begin{eqnarray}
\langle \zeta \rangle &\approx & 
\frac{1}{\frac{4}{3}\pi a^3}\int^{a}_{0} \zeta (z) dV
\nonumber \\
&\approx & \frac{1}{\frac{4}{3} \pi a^3}
\int^{r_2}_0 \frac{Z Z' e^2 \epsilon}{r} 4\pi r^2 dr
\end{eqnarray}
where $a$ is the mean
distance between atoms (typically
about $3 \times 10^{-8}\text{ cm}$ for a solid). 

Thus, for a solid, we estimate that
\begin{eqnarray}
\langle \zeta \rangle & \approx & \frac{3}{2} Z Z' e^2 \epsilon
\frac{r_2^2}{a^3}
\nonumber \\
&\approx & \epsilon \left( \frac{Z}{Z'} \right) 
10^{2} \ eV
\label{2}
\end{eqnarray}
Recall that the force due to the
electrostatic interactions depends on the
difference in $\langle \zeta \rangle$
between the two mediums [Eq.(\ref{1})].
This will be medium dependent, depending
on the chemical composition and structure
of the mediums. But, from Eq.(\ref{2}),
it is clear that the difference in $\zeta$ 
between two mediums will have the form:
\begin{eqnarray}
\langle \zeta_1 \rangle - \langle \zeta_2 \rangle =
\epsilon\lambda_{1,2} 10^2 \ eV
\label{3}
\end{eqnarray}
where $\lambda_{1,2}$ is the `medium dependent' part,
which is a number of order 1. In going from a low
density medium (such as air) to a high density
medium, solid earth, $\lambda_{1,2}$ is negative,
which implies an attractive force if $\epsilon$ is
negative and a repulsive force if $\epsilon$ is
positive. 

The force on the mirror fragment due to the
electrostatic interactions can be obtained
by combining equations , Eq.(\ref{1}) and Eq.(\ref{3})
\footnote{
Unless otherwise stated, we use natural units with $\hbar = c = 1$.},
\begin{eqnarray}
F_{\text{static}} &= &\frac{ (\gem{\zeta_1} - \gem{\zeta_2}) \rho' R^2}{M_{A'}}\hat{n} 
\nonumber \\
&=&
\epsilon \lambda_{1,2} 10^{13}\haak{\rho'/\haak{\text{g}/\text{cm}^{3}}}
(R/\text{cm})^2 \hat{n}\ \text{ g cm}
/\text{s}^2 
\label{helr}
\end{eqnarray}
where we have taken
$M_{A'} \sim 20 M_{\text{proton}}$, $A\sim R^{2}$ where $R$ is the size of the object. 

To find out if a mirror matter grain can remain at the Earth's surface, we have to compare this with the gravitational force $F_{\text{gravity}}$.
With our notation,
\begin{equation}
\lhaak{F_{\text{gravity}}}\sim \rho' R^3 g. 
\end{equation}
Hence,
\begin{equation}\label{stab}
\frac{\lhaak{F_{\text{static}}}}{\lhaak{F_{\text{gravity}}}}
\sim \lhaak{\epsilon} 10^{10} (\text{cm}/R)
\end{equation}
where we have used that $\lhaak{\lambda_{1,2}}\sim 1$ 

Recall that $F_{\text{static}}$ is 
the force on a mirror matter
fragment embedded in an ordinary
matter medium, where the fragment was at rest relative
to the medium. As discussed in Eq.(\ref{aa1}),
for a fragment moving with relative velocity $U$
there will be a velocity dependent frictional term, $F_{\text{friction}}$ as well.
Our purpose now is to estimate $F_{\text{friction}}$. The frictional
effect of mirror matter moving through an ordinary matter medium
has been considered previously in Ref.\cite{footyoon,eros}, but
in that case only the high velocity regime was examined 
($U \gg 1 \text{ km/s}$).
For the purposes of this paper, we are particularly interested in
the case where $U \lesssim 1 \text{ km/s}$, which hasn't
been evaluated previously. 

A mirror matter fragment moving 
through a homogeneous ordinary matter medium will experience 
a friction force caused by 
momentum transfer from collisions of mirror atoms with the ordinary 
atoms in the medium. If $U \lesssim v_{\text{thermal}}$
then the frequency of collisions suffered by a 
mirror atom is roughly $n v_{\text{thermal}}\sigma$, 
with $n = \rho/M_{A} \sim 10^{23}/\text{cm}^{3}$ is the number density 
of atoms in the 
ordinary medium, $v_{\text{thermal}}\sim\sqrt{6 k_{b}
T_{\text{room}}/m_{\text{atom}}}$ is the average relative speed of 
mirror atoms relative to the ordinary atoms, both assumed 
to be of mass $m_{\text{atom}}$, and $\sigma$ is the 
elastic cross section. In the Born approximation,
the differential cross section is given by\cite{mer}
\begin{eqnarray}
\frac{d\sigma}{d\Omega} = \frac{4 M_A^2 \epsilon^2 e^4 Z^2 Z'^2}
{(4M_A^2 U^2 \sin^2 \frac{\theta_{\text{scatt}}}{2} + \frac{1}{
r_2^2})^2}.
\label{rr3}
\end{eqnarray}
This is just the Rutherford formula cutoff at a distance $r_2$,
[Eq.(\ref{bohr})],
which is the range of the potential.
At low velocities, 
$U \stackrel{<}{\sim} 300$ m/s, 
the second term in the denominator dominates
over the first term and the cross section becomes approximately
isotropic, and Eq.(\ref{rr3}) reduces to
\begin{equation}
\sigma = 16\pi M_A^2 \epsilon^2 e^4 Z^2 Z'^2 r_2^4.
\end{equation}
Observe that for $|\epsilon | \stackrel{>}{\sim} 10^{-8}$,
$\sigma \stackrel{>}{\sim} r_2^2$ which is unphysical.
In fact, the interaction has become so
strong that the Born approximation breaks down.
For $|\epsilon | \stackrel{>}{\sim} 10^{-8}$, the
cross section saturates at $\sigma \sim r_2^2$. Thus, we have:
\begin{eqnarray}
\sigma &\sim & 10^{-2} \epsilon^2 \ \text{cm}^2 \text{ for } 
|\epsilon |
\lesssim 10^{-8}, \nonumber \\
\sigma &\sim & 10^{-18} \ \text{cm}^2 \text{ for } 
|\epsilon | \gtrsim 10^{-8}. 
\end{eqnarray}
Note that the above cross section is only
valid provided that $U \lesssim 300 
\ \text{m/s}$.
For larger velocities the cross section is suppressed by the 
first term in the denominator of Eq.(\ref{rr3}), see
Ref.\cite{footyoon,eros} for more discussion about the
high velocity regime. 

In a collision part of the relative momentum will be transferred. 
If the whole mirror matter fragment is moving with velocity $U$ 
relative to the medium, the momentum transferred by the collisions 
will average out to about $m_{\text{atom}}U$ per mirror atom per collision
\footnote{
Note that this implicitly assumes that the ordinary matter medium
is in the solid state. For a gaseous ordinary matter
medium such as air, air molecules would build up within the body and
move along with it, which would effectively reduce the size
of the frictional force compared to a solid (even taking account
of the density difference). For a gaseous medium
it is better to work in the rest frame of the mirror body, and
examine the momentum transferred by the impacting air molecules
(as was done in Ref.\cite{footyoon}).
}. 
Therefore the friction force $F_{\text{friction}}$ exerted on 
the fragment of mass M 
(taking $m_{\text{atom}}\sim 20 M_{\text{proton}}$)
is approximately:
\begin{equation}\label{friction}
F_{\text{friction}}\sim 
M\sqrt{\frac{6 k_{b}T_{\text{room}}}{m_{\text{atom}}}}n
\sigma U
\sim 10^{9}
\Lambda
\haak{M/\text{g}}\haak{U/\haak{\text{m/s}}}
\left( \frac{\rho}{\text{g}/\text{cm}^3}\right) 
\text{ g m}/\text{s}^{2} 
\end{equation}
where $\Lambda =
\haak{|\epsilon |/10^{-8}}^{2}
$ for $|\epsilon | \lesssim 10^{-8}$ and $\Lambda = 1$ for
$|\epsilon | \gtrsim 10^{-8}$. 

From Eq.\eqref{friction} we find that a mirror fragment
with initial velocity $U_i$
will slow down enough to enable it to be captured in
ordinary matter after a distance of order:
\begin{eqnarray}
L \sim 
\frac{10^{-7}}{\Lambda}\haak{\frac{U_i}{300\text{ m/s}}}\left(
\frac{4 \text{ g}/\text{cm}^3}{\rho}\right)
\ \text{ meters}
\label{fa}
\end{eqnarray}
Recall that the above equation is roughly valid for $U_i \lesssim 
300$ m/s. For completeness, let us mention that in the case of 
$U_i \gtrsim 300$ m/s, the corresponding
distance is\cite{footyoon,eros}:
\begin{equation}
L \sim \frac{U^4_i M^2_{A'}M_{A}}{160\pi \rho Z^2 Z'^2 \epsilon^2
e^4}
\sim 10^{-7}\left( \frac{U_i}{300 \text{ m/s}} \right)^4 \left( \frac{10^{-8}}{
\epsilon}\right)^2 
\left( \frac{4 \text{ g}/\text{cm}^3}{\rho}\right) 
\text{ meters.}
\label{tye}
\end{equation} 
Anyway, the net effect is that for low velocities,
$U \lesssim 1 \text{ km/s}$
we see from Eq.(\ref{fa},\ref{tye}) that mirror matter fragments
rapidly slow down in ordinary matter. 
 
\end{document}